\documentclass[10pt]{iopart}
\usepackage{graphicx}
\usepackage{natbib}
\usepackage{amssymb}
\bibpunct{(}{)}{;}{a}{}{,}

\begin{document}

\review[Neutron-capture abundances in metal-poor stars]{Observational nuclear
astrophysics: Neutron-capture element abundances in old, metal-poor stars}

\author{Heather R.\ Jacobson and Anna Frebel}

\address{Kavli Institute for Astrophysics and Space Research and\\
Department of Physics \\ Massachusetts Institute of Technology \\ 77
Massachusetts Avenue, Cambridge, MA 02139, USA}
\ead{\mailto{hrj@mit.edu}, \mailto{afrebel@mit.edu}} 

\begin{abstract}
The chemical abundances of metal-poor
stars provide a great deal of information regarding the individual
nucleosynthetic processes that created the observed elements and the overall process
of chemical enrichment of the galaxy since the formation of the first
stars.  Here we review the abundance patterns of the neutron-capture
elements ($Z \ge 38$) in those metal-poor stars and our current understanding of 
the conditions and sites of their production at early times.  We also review 
the relative contributions of these different processes to the 
build-up of these elements within the galaxy over time, and
outline outstanding questions and uncertainties that complicate
the interpretation of the abundance patterns observed in
metal-poor stars.  It is anticipated that future observations of large 
samples of metal-poor stars will help discriminate between 
different proposed neutron-capture element production sites
and better trace the chemical
evolution of the galaxy.
\end{abstract}


\section{Introduction}\label{intro}
\subsection{Spectroscopy in Astronomy} 
The ability to perform spectroscopy on starlight transformed the
science of astronomy and the study of cosmic objects.  For millenia,
astronomers mapped the positions of the stars, traced the orbits of
the planets, and recorded the appearance of novae and other phenomena
in the sky.  But that was as far as they could go: questions such as
how far away stars were (parallax being of extremely limited use), and
even more fundamentally, what they were made of remained unanswerable
until starlight was passed through a telescope and into a
spectrograph.  It can also be said that spectroscopy first connected
astronomy to the sciences of chemistry and physics in profound ways.
Whereas Newton united the heavens and Earth with universal laws of
motion, studying the spectra of stars helped unravel the secrets of
the atom, identified the processes in which elements in the Periodic
Table formed, and how the universe became the one we live in now.

Spectroscopy makes this possible by allowing us to see absorption and
emission 
features in starlight as it is spread out by a prism or a grating to
form a spectrum. These absorption features are caused by atoms and
molecules in the star's atmosphere absorbing photons coming from the
star's interior at discrete energies (wavelengths/frequencies).  As
demonstrated by Fraunhofer, Kirchhoff, Bunsen and others in the 19th
century, each element or molecule interacts with photons in a unique
range of wavelengths, giving each a unique chemical fingerprint.  It
is therefore possible to identify a species of atom or molecule in a
stellar atmosphere based on the presence of its absorption line(s);
furthermore, it is possible to infer its abundance (by number) based
on the strength of its absorption line(s).  Comparing the relative
amounts of different chemical species in stars of different ages and
locations allows astronomers to trace the history of the production of
chemical elements in the universe.

\subsection{Chemical Evolution}
\begin{figure}[!ht]
  \begin{center}
     \includegraphics[width=14cm]{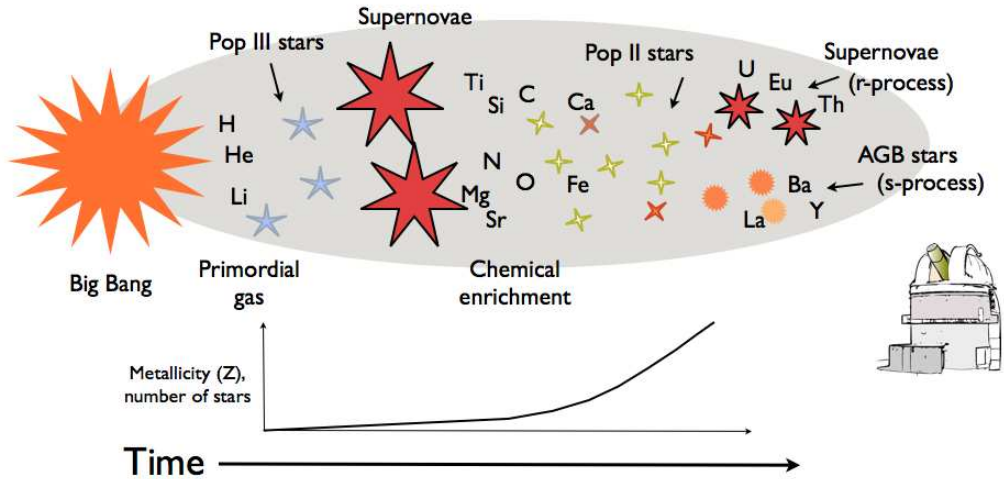}
     \caption{ Simple illustration of chemical enrichment of the
       universe: massive Population\,III stars form out of primordial gas,
       explode as supernovae, and enrich the interstellar medium with
       products of stellar nucleosynthesis.  Subsequent cycles of star
       formation and death (Population\,II) steadily enrich the universe with metals
       over time. The first low-mass stars to form in the universe
       are still observable today.  Two main contributors to chemical
       enrichment after the first stars are $8-10$\,M$_{\odot}$ stars
       that explode as core-collapse supernovae, and less massive stars that
       enrich the interstellar medium via strong mass loss and stellar
       winds (AGB stars).  Their nucleosynthetic products, the
       r-process and s-process elements, are the subject of this
       review.
      \label{cartoon}}
    \end{center}     
   \end{figure}

According to the Big Bang theory, the cosmic fireball that birthed our
universe created hydrogen, helium and a fine dusting of lithium.  All
the other elements in the Periodic Table (typically referred to as
``metals" in astronomy) were forged by nuclear reactions within
stars, or else in their supernova explosions.  Different metals are
produced by specific chains of reactions that occur at different
ranges of temperatures and densities, and in some cases are controlled
by the number of seed nuclei needed for the reactions.  Chemical
evolution is therefore a product of stellar evolution, as different
nucleosynthetic processes turn on and off as stars evolve (see, e.g.,
the seminal work of \citealt{b2fh}).  Figure~\ref{cartoon} shows a
simple schematic of this process.

Our Sun, with an age of 4.6\,Gyr, reflects the effects of some 8\,Gyr of
chemical evolution.  Many generations of stars enriched the gas from
which the Sun and the solar system formed, combining to produce the
amount of iron (for example) in our Sun's atmosphere, in the Earth's
core, and in our blood.  Such ``layers'' of chemical evolution are
difficult to disentangle, making it very difficult to directly compare
the chemical yields from theoretical supernova models to a star like
our Sun.  However, as chemical enrichment is a product of successive
cycles of star formation and evolution, it is possible to trace the
chemical history of our galaxy, and even to study single episodes of
chemical enrichment, by studying the oldest stars.

As Figure~\ref{cartoon} illustrates, the first stars to form in the
universe (so-called Population\,III stars for historical reasons)
formed from pristine clouds of H and He (with maybe tiny amounts of
Li), and after their short lifetimes\footnote{Current theory of star
  formation in the early universe predicts Population\,III stars to
  have been quite massive (10 to 100\,$M_{\odot}$), and therefore all
  are expected to have exploded as supernovae  a few
  million years after they formed \citep{brommARAA}.}, they polluted
interstellar gas with the products of nucleosynthesis in their
interiors and in their supernovae.  The subsequent generations
of stars that formed from this enriched material 
(correspondingly called Population\,II) inherited the chemical imprint of the first
generation, and then further enriched the interstellar medium with
products of their nucleosynthesis in the late stages of their
evolution (supernovae, or AGB stars, discussed in more detail later).  It is
expected that the low-mass stars from this second generation, which
were able to form for the first time, still shine in the universe
today given their long lifetimes ($>10$\,Gyr).

\subsection{Metal-Poor Stars}
It is now worthwhile to more specifically characterize the early,
second-generation Population\,II stars discussed in the previous
section.  Generically, they are called ``metal-poor" stars, to
indicate the relative paucity of the products of stellar
nucleosynthesis in their atmospheres, compared to that of the Sun,
which almost always serves as the reference.  Iron, Fe, is typically
used as a proxy for metallicity
because the large number of Fe absorption lines present in the optical
wavelength regime makes it straightforward to measure.  A prefix is
often used to illustrate how metal-poor a star is: ``extremely
metal-poor'' ($\mbox{[Fe/H]} < -3$\footnote{Astronomers employ the
  [A/B] notation to describe the relative abundances of two elements
  in a star compared to that in the Sun: [A/B] =
  $\log_{10}$(N$_{A}$/N$_{B}$) $-$
  $\log_{10}$(N$_{A}$/N$_{B}$)$_{\odot}$.  A star with $\mbox{[Fe/H]}
  = -2$, for example, contains a factor 100 fewer Fe atoms by number
  than the Sun.}), ``ultra metal-poor" ($\mbox{[Fe/H]} < -4$), and
``hyper metal-poor" ($\mbox{[Fe/H]} < -5$)\citep{ARAA}.  The detailed
element abundances of these stars are used to reconstruct the physical
and chemical processes of early star and galaxy formation and
constrain our understanding of the early universe.

Major topics include:
\begin{itemize}
\item The origin and evolution of the chemical elements
\item The relevant nucleosynthesis processes and sites of chemical element production
\item The nature of the first stars and their initial mass function
\item Early star and galaxy formation processes
\item Nucleosynthesis and chemical yields of the first/early supernovae
\item The chemical and dynamical history of the Milky Way
\item A lower limit to the age of the universe
\end{itemize}

The sixth item requires large samples of stars covering a wide range
of age and metallicity (see Section~\ref{ncaps}), while the last is
made possible by the detection of radioactive elements such as Th and
U in selected, individual stars (see Section~\ref{chrono}).

As alluded to in the previous section, very different physical
conditions are responsible for the production of elements in different
regions of the Periodic Table.  Studies of metal-poor
stars have shown that the production of the light elements (Fe-peak
and lighter, atomic number $Z\leq30$) is decoupled from that of the
heavier elements.  For example, trends of e.g., [Mg/Fe] versus [Fe/H]
for metal-poor stars shows a very small scatter of order 0.1 dex,
while the [Ba/Fe] ratios versus [Fe/H] for the same stars have a
scatter of $>$1 dex \citep{heresII}.  
For the remainder of this review, we focus mainly on what metal-poor
stars reveal about the production of the neutron-capture elements
($Z>38$) and their production sites in the early universe.  
An extensive earlier review on neutron capture element abundances
can also be found in \citet{sneden_araa}.
For a review of the lighter elements, we refer
the reader to \citet{frebel_bookchapter}.

\subsection{How Metal-Poor Stars are Found}
First though, a brief summary of how metal-poor stars are found and
studied is necessary.  Metal-poor stars are extremely rare objects
which makes finding them a great challenge. Techniques are required to
efficiently sift through the large numbers of younger, metal-rich
stars to uncover fewer than 0.1\% of survivor stars from the early
universe. Large-scale systematic searches began with the HK survey by
\citet{BPSI,BPSII} which were then superceeded by the Hamburg/ESO
Survey \citep{hespaperI, hes4}. The HES covered $\sim$1000 square
degrees of the southern sky collecting data of some 4 million point
sources.

\begin{figure}[!ht]
  \begin{center}
     \includegraphics[width=10cm]{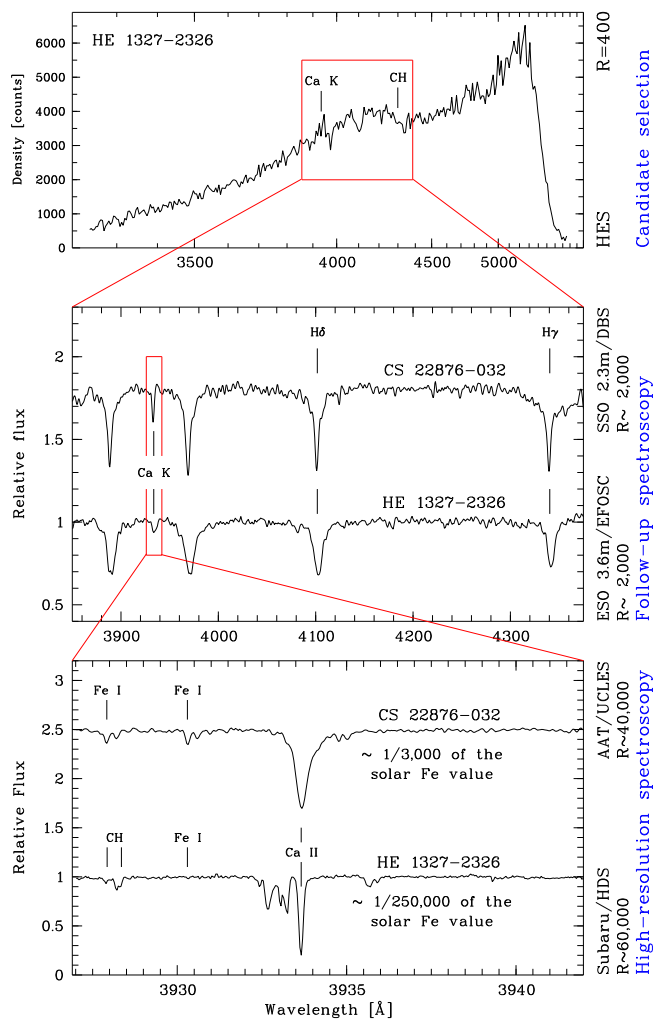}
     \caption{The process of finding a metal-poor star.  Stars with weak
     Ca\,II\,K lines are identified in low-resolution spectra
     (top), and are selected for follow-up with medium-resolution
     spectroscopy (middle) to get a direct measure of the Ca\,II\,K line
     strength.  The most metal-poor stars of this sample are then
     selected for high resolution followup (bottom), where the abundances
     of other elements can be determined.  Figure taken from \citet{Frebel2005IAU}.
     \label{spec}}
    \end{center}     
   \end{figure}

The low-resolution ($R=\lambda/\Delta \lambda\sim15$\,{\AA})
objective-prism spectra collected in both surveys cover the strong
resonance absorption line of calcium, the Fraunhofer ``K'' line,
located at 3933\,{\AA} which can be used as a metallicity
indicator. Stars that show a weak Ca\,K line as a function of the
surface temperature (temperature affects the line strengths) are
selected as candidate metal-poor stars. It is usually assumed that the
calcium abundance traces the overall metallicity. To confirm a stars'
low-metallicity nature, additional spectra with higher resolution are
required. Those have $R\sim2000$, are usually obtained with telescopes
with 1 to 4\,m mirrors, and Figure~\ref{spec} shows examples. These
spectra allow a much more refined measurement of the Ca\,K line
strength.  Together with the color of the star, they can be turned
into a metallicity either via line-strength-color-calibrations
\citep{BeersCakII} or through fitting large portions of the spectrum
with grids of synthetic spectra of known temperature and metallicity
(e.g., \citealt{lee_sspp1}).

Once the low-metallicity nature is confirmed from medium-resolution
spectra, high-resolution spectra are required for a detailed abundance
analysis of many elements, including iron. Spectral lines of Fe and
other elements (e.g., Mg, Si, Ti) are
very weak and detectable in spectra with $R>20000$ which are
obtained with high-resolution spectrographs on typical
$\sim$6 to 10\,m telescopes.

More recent searches employ slightly modified approaches, such as
surveying the sky immediately with medium-resolution spectroscopy
using large multi-object spectrographs (the Sloan Digital Sky Survey
and its SEGUE follow-up survey, and the LAMOST survey) or selecting
metal-poor candidates from photometric survey data. The SkyMapper
telescope is photometrically surveying the southern sky in specific
filter sets (e.g., \citealt{bessell11}) that allow candidate
selection in a very efficient way. Follow-up with medium- and
high-resolution spectroscopy is still required, though.

\section{Neutron-Capture Nucleosynthesis}
Elements in the periodic table beyond the iron-peak are primarily formed via the
capture of neutrons on to seed nuclei such as iron.  This process can
occur on two timescales.  In the slow neutron-capture process
(s-process), the rate of capture of neutrons on to seed nuclei is slow
enough to allow the unstable nuclei to $\beta$-decay to stable nuclei
before subsequent capture.  In the rapid neutron-capture process
(r-process), the rate of capture is much greater than the rate of
$\beta$-decay, resulting in the build-up of heavy unstable
neutron-rich nuclei that then decay to form heavy, stable nuclei along
the ``valley of $\beta$ stability" \citep{sneden_araa}.  Each process
produced roughly 50\%\ of all the neutron-capture isotopes in the
solar system \citep{arlandini1999}. However, given their different
timescales, it is believed that the majority of all elements with
$Z\geq38$ in the early universe (first few stellar generations) were
formed via the r-process, with chemical enrichment from low-mass AGB
stars coming at later times (e.g., \citealt{argast}).  A significant
contribution to the production of elements with $Z \gtrsim 38$ is also
predicted to come from massive ($>8\,M_{\odot}$) stars, either by
charged particle reactions in core collapse supernovae and/or by the
s-process (e.g., \citealt{Pignatari2010}).
 In this Section we begin to lay out the nucleosynthesis
details of the various neutron-capture processes to then tie them to
observations in metal-poor for an understanding of chemical evolution
in the next sections.

\subsection{The s-Process nucleosynthesis}\label{sproc_nucleo}
Detailed reviews of the development of s-process nucleosynthesis
theory and observations can
be found in, e.g., \citet{busso_gallino_AGB1999}, \citet{busso2001},
\citet{gallino1998}, and \citet{sneden_araa}.  Given that much of the
s-process evolves along the valley of $\beta$ stability, most of the neutron
capture rates involved can be investigated in the laboratory.  As a result,
the s-process is qualitatively well-understood and the s-process
abundance pattern for the Sun can be well reproduced, especially for
$Z > 56$, considering also information from calculations of galactic
chemical evolution (e.g., \citealt{Cameron1973},
\citealt{2000burris}, \citealt{travaglio}).
The solar s-process pattern, which serves as the reference to which
other stars are compared, shows three distinct peaks in the abundance
distribution: the first peak at $Z=38-40$ (Sr, Y, Zr), the second peak
at $Z=56-60$ (Ba through Nd), and the third peak at $Z=82-83$, the
nuclei at which the s-process ends (Pb, Bi).  Elements in each of
these peaks are produced by different neutron exposures 
\citep{busso_gallino_AGB1999, sneden_araa}.

Observations and theory agree that the so-called ``main'' s-process
operates in low- and intermediate mass stars ($\sim1-8$\,M$_{\odot}$;
\citealt{busso_gallino_AGB1999}) in the last $\sim1\%$ of their
lifetime as they evolve along the asymptotic giant branch (AGB) in the
Hertzsprung-Russell diagram\footnote{The first sign that neutron
  capture nucleosynthesis occurs in stars was the detection of the
  unstable element Tc in an evolved star by \citet{merrill_tc}.}.
During the AGB phase of stellar evolution, a star has an inert carbon
oxygen core successively layered with a helium-burning shell, a
helium-rich region, a hydrogen-burning shell, and then a convective
envelope (see, e.g., Herwig 2005 for a review). 

In the classical picture, more than 90\% of the neutrons are formed via the $^{13}$C($\alpha$,
n)$^{16}$O reaction in between thermal pulses that occur in AGB stars
\citep{straniero1995}.
But recent work has indicated that this scenario is likely
simplistic and more neutron-capture regimes may operate, especially at low metallicity
(see \citealt{lugaro2012}).
For this reaction to be activated, a radiative $^{13}$C pocket must form
after protons from the envelope are mixed down into the intershell
layer in order to combine with $^{12}$C to form $^{13}$C by partial
completion of the CN cycle.  It is then
that the s-process operates. A subsequent convective thermal pulse
then mixes the s-process products throughout the He intershell region
(e.g., \citealt{straniero1995}, \citealt{gallino1998},
\citealt{herwig_araa}).  Repeated dredge-up processes finally mix the
material from the inner regions of the star to the surface. This
sequence of processes happens with each thermal pulse which are on order
$10^{4}$ to $10^{5}$ years apart. 

Several physical mechanisms have been proposed to explain the
formation of the $^{13}$C pocket (see \citealt{herwig_araa} for a
review).  Nevertheless, the cause of this mixing remains unclear, and
several prescriptions have been developed to simulate the formation of
the $^{13}$C pocket and the s-process production that occurs in it
(e.g.,\citealt{gallino1998}, \citealt{goriely&mowlavi2000}, 
\citealt{herwigetal:2003}, \citealt{bisterzo2011}, \citealt{maiorca2012},
\citealt{lugaro2012}).

The remaining 10\%\ of neutrons formed in AGB stars are created via
the $^{22}$Ne($\alpha$, n)$^{25}$Mg reaction during convective thermal
pulses.  This reaction results in a higher neutron density compared to
that in the $^{13}$C pocket ($n_{n} > 10^{10}$ cm$^{-3}$,
\citealt{herwig_araa} and references therein, \citealt{lugaro2012},
\citealt{karakasetal:2012}, \citealt{vanraietal:2012}).  The contribution of
this process to the s-process element abundance distribution is
smaller than that of the $^{13}$C($\alpha$, n)$^{16}$O reaction.
However, the isotopic distribution is greatly affected because of the
activation of several branching points along the s-process path. Using
the resulting abundance pattern, the conditions in the AGB He
intershell can be studied in great detail (e.g., \citealt{kaeppeleretal:2011}). 

\begin{figure}[!hb]
  \begin{center}
    \includegraphics[width=10cm,
      angle=0]{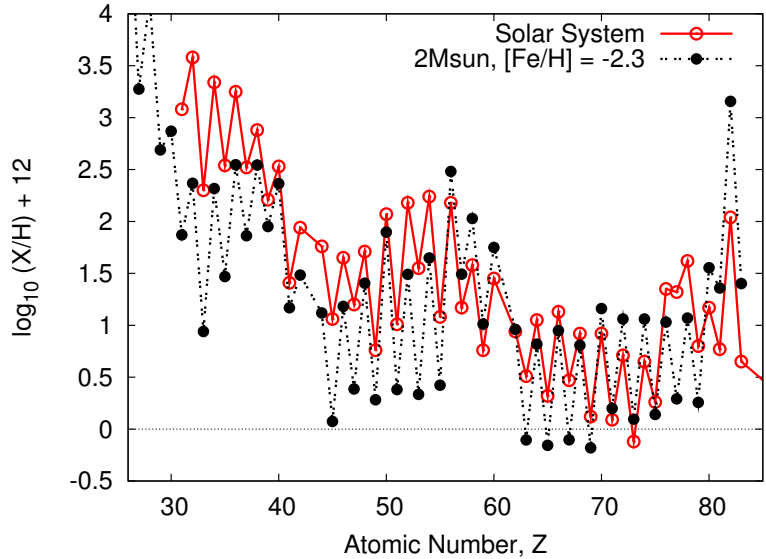}\caption{ The neutron-
      capture element abundance 
      pattern of the Sun compared to that of a 2$\,M_{\odot}$ model
      AGB star with $\mbox{[Fe/H]}=-2.3$ \citep{lugaro2012}.
      Despite the difference in
      metallicity, the s-process operates efficiently enough in this
      metal-poor star to produce s-process abundances equal or greater
      than the solar system values.  Note the enhanced production of
      lead (Pb, $Z=82$). Figure from A.\ Karakas, used with permission.
       \label{sproc_pattern}}
    \end{center}     
   \end{figure}

The s-process efficiency depends on its seed nuclei, therefore the
abundance pattern is different in stars of different metallicity
(e.g., \citealt{gallino1998}, \citealt{busso2001}).  
Figure~\ref{sproc_pattern} illustrates
this: here, the solar system s-process pattern is compared to that for
a model of a low-metallicity AGB star.  The low-metallicity AGB star
has a larger ratio of neutrons to seed nuclei compared to a
solar-metallicity star, and therefore the s-process preferentially
produces heavier species. It still results in good agreement between
the two patterns for the heavier ($A \gtrsim 170$) species, but larger
discrepancies for the lighter species ($90 \le A \le 130$)
\citep{karakas2010, lugaro2012}. Note also the much larger Pb abundance of the
low-metallicity star (see also Section~\ref{sproc_abund}).

The light s-process elements up to the first peak (which are formed in
the ``main'' s-process in low-mass AGB stars) can also be made by the
s-process in intermediate-mass ($>$3\,M$_{\odot}$) AGB stars or in
massive ($>$8\,M$_{\odot}$) stars during core He and
shell C burning (e.g., \citealt{lamb1977}, \citealt{bussogallino1985},
\citealt{the2007}, \citealt{Pignatari2010}).  Here, neutrons are
created via the $^{22}$Ne($\alpha$, n)$^{25}$Mg reaction.  This
process generates higher neutron fluxes on shorter timescales than the
main s-process described above (e.g., \citealt{gallino1998}).  Given
the shorter lives of massive stars relative to low mass stars, it is
possible that massive stars contributed some of the s-process element
enrichment in the universe earlier than the main s-process.

Indeed, models of massive, metal-poor and fast-rotating
($\sim500-800$\,km\,s$^{-1}$) stars indicate that rotation-induced
mixing within the star leads to production of large amounts of
$^{14}$N, $^{13}$C and $^{22}$Ne, the latter of which provides
neutrons for the s-process \citep{pignatari, frischknecht2012}.  Fast
rotation can boost s-process element production by orders of magnitude
(e.g. for Sr; \citealt{chiappini11} and references therein).  Chemical
enrichment from these stars will be discussed further in
Section~\ref{sproc}.

\subsection{The r-Process nucleosynthesis}\label{rproc_nucleo}

Numerous sites for the r-process have been proposed over the years and
generally fall into two broad categories: neutron star - neutron star
or neutron star - black hole mergers and supernovae.  The merger of
compact objects can easily provide the neutron flux needed for rapid
neutron-capture to occur (e.g., \citealt{lattimer_schramm74,
  lattimer_schramm76}; \citealt{freiburghaus99};
\citealt{goriely2011}).  However, as pointed out by
\citet{argast2004}, the timescales and relative rarity of such events
are such that they cannot account for the existence of extremely
metal-poor stars exhibiting an r-process signature that presumably
formed long before the first neutron-star mergers occured.  Therefore,
while they could contribute to some r-process enrichment, neutron star
mergers are unlikely to be the main site in the early universe.

Assuming that the r-process takes place during core collapse of
massive stars ($8 - 10\, M_{\odot}$; e.g., \citealt{wanajoetal:2003}),
their explosion mechanisms and properties have also been explored in
recent years, but each has its own drawbacks, such as failing to
provide sufficient explosion energies, entropies or neutron fluxes to
drive the r-process (see, e.g., \citealt{ct2004, arnould2007,
  sneden_araa} and references therein).  That said, researchers have
nonetheless explored the range of possible neutron-capture reactions
using the so-called ``waiting point method" that is independent of the
explosion site.  Here, the details of the explosion are not
considered, but rather its energy output is used to explore the
parameter space of neutron density, electron abundance, neutron flux,
and entropy that can produce an abundance pattern similar to that seen
in the Sun as well as metal-poor stars.  Some of these models are
called the ``neutrino wind model" and the ``high entropy wind model",
to name two (e.g., \citealt{woosley_hoffman1992, wanajo01}).  These
studies have found that the neutron densities and entropies required
to produce the light neutron-capture species (e.g., Sr, Y, Ba,
Z$\approx$38-40) differ by orders of magnitude from those needed to
produce the heavier isotopes (the lanthanides and actinides;
\citealt{montes, arcones_montes2011, kratz07, farouqi2009}).
While many advances have been made over the last $\sim$50 years,
the r-process calculations remain a challenging task given the
difficulty to obtain experimental data of the most neutron-rich
nuclei and the uncertainties relating to the astrophysical site.
 
\section{Neutron-capture element abundances in metal-poor stars}

A small subset of metal-poor stars show a strong enhancement of
neutron-capture elements compared to iron and lighter elements with
$Z\le30$. The task at hand then is to identify which nucleosynthesis
process was responsible for the creation of these elements that are
now observed in those metal-poor stars. The relative enrichment of
metal-poor stars by the r- and s-processes compared to Fe can be
distinguished by comparing the abundances of elements predominantly
produced by either process, or by comparing an element that may be
produced by both processes to that produced only in one.  For example,
lead (Pb) is produced mainly by the s-process, and similarly
barium. On the contrary, europium is mainly made by the r-process
(e.g., \citealt{simmerer2004}).  Stars enriched only by the r-process
can therefore be identified by their [Pb/Eu] or [Pb/Ba] ratios (e.g.,
\citealt{roederer_ubiq}) or [Ba/Eu] ratios (e.g., \citealt{heresII}).

\subsection{s-Process element abundances}\label{sproc_abund}
The time-scale over which s-process enrichment occurs after the
formation of the first generations of stars is delayed by up to a
billion years due to the long main-sequence lifetimes of the first
low-mass stars before passing through the AGB phase.  $10-20\%$ of
metal-poor stars in the halo (which can have ages up to $10-12$\,Gyr)
display large enhancements of s-process elements \citep{cohen2006,
  lucatello2006}. The best explanation for the existence of such stars
is that their atmospheres were polluted by a slightly more massive
binary companion that passed through the AGB phase and transferred
s-process material on to them (along with large quantities of carbon,
another nucleosynthetic product dredged up to the surface of a star
during AGB evolution; \citealt{sneden_araa, lugaro2012, placco2013}).  
Radial
velocity studies of such stars indeed show the majority of them to
move around a by now unseen companion \citep{lucatello2006}.

A prominent feature of s-process enriched metal-poor stars is that
they show high abundances of the heaviest s-process element, Pb
(e.g., \citealt{vaneck2001}, \citealt{ivans05}, \citealt{cohen2006}, \citealt{placco2013}).
Lead is the end-product when the s-process is allowed to run to
completion.  The large Pb abundances of some metal-poor stars are
consistent with the theory that Pb is produced in large quantities when
the ratio of neutrons to seed nuclei is high (e.g.,
\citealt{gallino1998}), a condition easily met in metal-poor stars 
(Section~\ref{sproc_nucleo}; Figure~\ref{sproc_pattern}).
\citet{aoki_lead2002} observed s-process element enriched stars that
showed a large scatter ($>$1\,dex) in [Pb/Ba] ratios, greater than
that predicted by low-metallicity AGB model yields.  Such scatter in
s-process element abundances in low-metallicity environments provide
useful constraints on AGB stellar evolution models \citep{herwig_araa,
  lugaro2012}. Consequently, the s-process is not a universal process
but dependent on the metallicity of the star, and hence, its time of
formation in the universe.

For completeness, we note the existence of a small sub-class of metal-poor
stars exhibiting neutron-capture elements from both the s- and
the r-process.  It is further discussed in \citet{sneden_araa}.

\subsection{r-Process element abundances}\label{rproc_abund}

In contrast to the s-process, the time-scale for neutron-capture and
build-up of heavy isotopes via the r-process is of order seconds rather
than millennia.  The r-process contribution to the neutron-capture isotopes 
in the solar system is calculated by subtraction of the solar s-process
pattern from the total solar system isotopic abundances (as derived from
the Sun's atmosphere and meteorites).  The residuals are defined 
to be ``the'' r-process \citep{arlandini1999, bisterzo2011}.

\begin{figure}[!ht]
  \begin{center}
     \includegraphics[width=14cm]{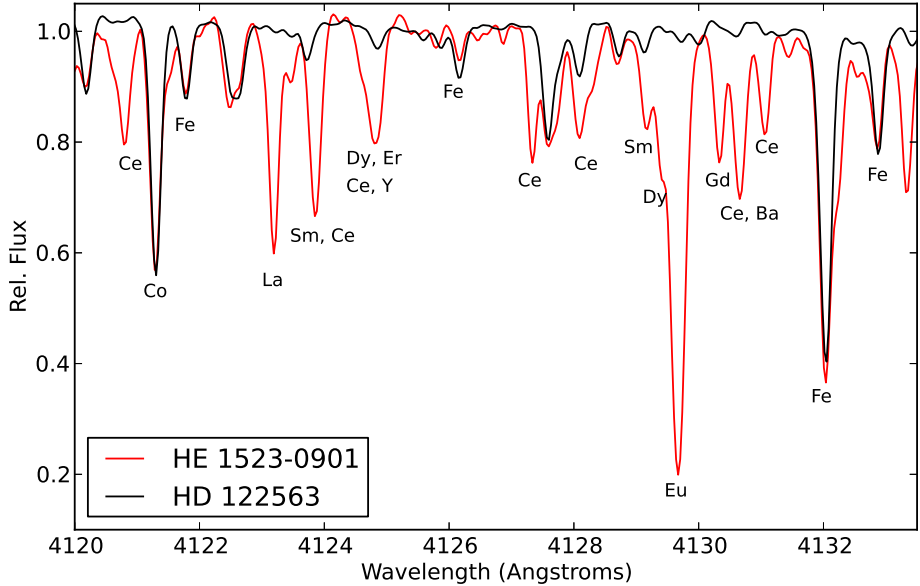}\caption{
      Portions of high resolution ($R=\lambda/\Delta \lambda\approx 30000$) 
      spectra of the r-process
      element rich star HE~1523$-$0901 (red) and the r-process
      element deficient star HD~122563 (black).  These two stars have
      similar [Fe/H] values and atmospheric parameters.
      Various element absorption
      lines are identified.\label{rprocspec}}
    \end{center}     
   \end{figure}

A small fraction of stars with $\mbox{[Fe/H]} <-2.5$ (e.g., 5\% of
giant stars; \citealt{heresII}) show unusually large
enhancements in r-process elements ($\mbox{[r/Fe]}>0$).
Figure~\ref{rprocspec} shows the spectra of two stars, one with such
large enhancements of neutron-capture element abundances, and another
without enhancements.
These r-process enriched stars provide important constraints on the
site(s) and mechanism(s) of the r-process because they likely formed
in the vicinity of a recent supernova event at the earliest times and
during which the r-process took place.

Consequently, two general scenarios have been proposed to explain the
origin of such large r-process element enhancements in metal-poor
stars. The first scenario proposes that they are simply the result of
inhomogeneous mixing of r-process material with the ISM, inheriting
larger quantities of neutron-capture species compared to others in the
same generation \citep{argast}.  The second scenario proposes a
``local" r-process enrichment due to a close binary companion,
\citep{qianwasserburg2001}.  Recently, \citet{hansen2011} presented a
test of the second scenario.  They carried out a long-term radial
velocity study of 17 r-process enhanced giant stars over a 4 year
period.  Of their sample, 14 had no detectable variation in RV,
indicating lack of binarity.  Therefore binarity and pollution by a
companion are unlikely to explain strong r-process enhancement for the
majority of stars.  However, the inhomogeneous mixing scenario has its
own complications (Section~\ref{rproc}).

\subsubsection{Universal ``Main'' r-process}\label{main_rproc}

\begin{figure}[!ht]
  \begin{center}
     \includegraphics[width=14cm]{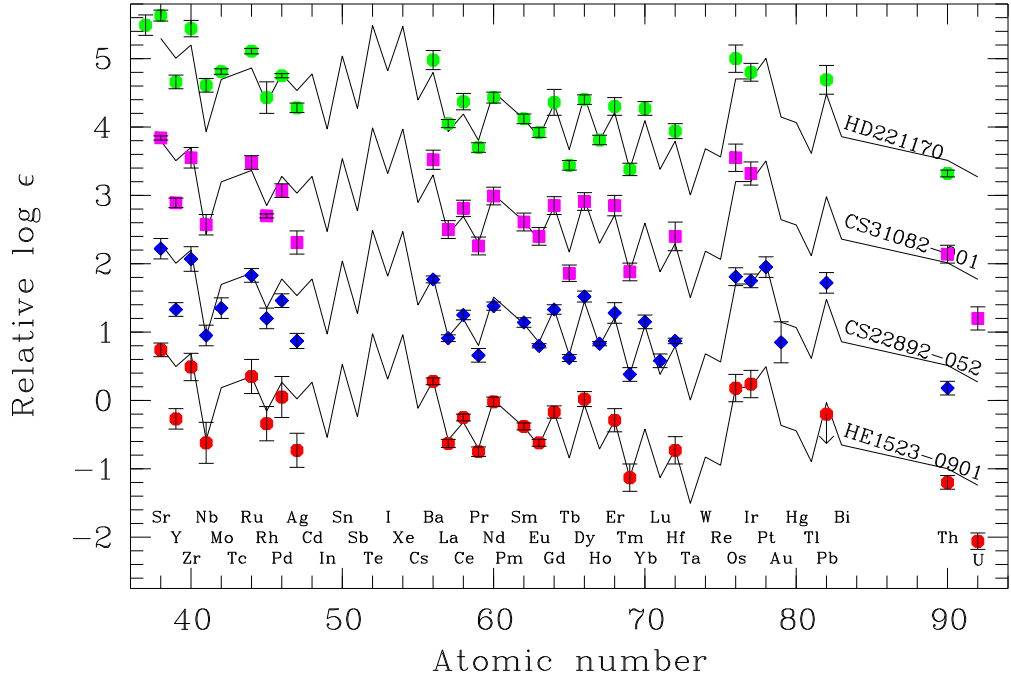}
     \caption{ Element abundance patterns of a sample of r-process
       element enriched stars (points), with the solar r-process
       pattern scaled to each star's Eu abundance (solid lines).  The
       agreement between the solar pattern and the stellar abundances
       is excellent for elements $Z\approx56-72$.  However, the
       agreement with the solar pattern of the lighter elements
       ($Z\approx38-40$) is not as good.  References for the
       abundances: HD~221170 -- \citet{ivans06}; CS~31082-001 --
       \citet{Hilletal:2002}; CS~22892-052 -- \citet{Snedenetal:1996};
       HE~1523$-$0901 -- \citet{he1523}.
       (Figure adapted from
       \citealt{frebel_bookchapter}.) \label{rpattern} }
    \end{center}     
   \end{figure}

Stars with unusually strong Eu absorption lines in their spectra are
now regularly identified in spectroscopic studies and can be shown to
contain large amounts of r-process elements. The first such star to be
discovered, CS 22892-052 (\citealt{Snedenetal:1996, sneden03}), 
has an element abundance distribution that
matches the scaled solar system r-process pattern remarkably well.  As
the number of r-process enhanced metal-poor stars discovered over the
years has grown, many more have been shown to follow the solar system
r-process pattern as well (e.g., \citealt{sneden_araa}).
Figure~\ref{rpattern} shows examples of this: the abundances of four
r-process element enriched stars are shown, along with the solar
r-process pattern (solid lines).  The agreement is excellent for
elements heavier than Ba.  This ``universality'' of the abundance
pattern in r-process enriched metal-poor stars and the Sun indicates
that the r-process mechanism essentially operates identically wherever
and whenever suitable conditions are present.  It is present in the very first
generations of stars and appears unchanged throughout the chemical
evolution that culminated in the formation of the Sun.

\subsubsection{Deviations from the scaled solar r-process pattern}\label{dev_rproc}

However, not all neutron-capture elements observed in r-process
metal-poor stars perfectly follow the scaled solar pattern. The light
neutron-capture elements up to barium show deviations in their
abundances, as can be seen in Figure~\ref{rpattern}. These deviations
are well documented by now but remain unexplained thus
far (e.g., \citealt{heresII, honda06, honda07, roederer_ubiq, hansen13_agpd, yong13_III}). 
Suggestions include whether these deviations from the scaled
solar pattern as well as the scatter found among the known r-process stars
reflect observational uncertainties. Many of these elements are
difficult to detect in stellar spectra. Also, an additional
nucleosynthesis process might be operating in this region producing light
neutron-capture elements in addition to the main r-process. This would
alter the overall abundance pattern in metal-poor stars compared to
that of the sun. Alternatively, there remains the possibility that the
production of elements lighter than barium simply is not universally
possible.

\subsubsection{More deviations: ``actinide boost''}\label{act_boost}

Another interesting, so far unexplained, phenomenon has appeared among
strongly r-process enhanced stars. About a quarter of this group of
objects shows thorium abundances that are higher than expected
compared to other stable r-process elemental abundances and the scaled
solar r-process pattern. This behavior has been termed ``actinide
boost'' \citep{honda04, lai2007, Hilletal:2002}, and most prominently results in
\textit{negative} stellar ages when using the Th/Eu chronometer 
(see also Section~\ref{chrono}) since the
decay of thorium has not been lasting since the time of the star's
formation. One explanation may be that these stars show the r-process
pattern of two r-process events that occured at different times -- one
just prior to the star's formation and one at a later time in the
vicinity of the star. This way, the {\it pattern} of stable r-process
elements would be preserved (albeit not the overall amount), but the
radioactive and thus decaying element abundances would be higher than
in the case of just the initial r-process abundance level the star was
born with. While this explanation is qualitatively straight forward,
it remains to be seen how a star could realistically acquire any
material from such a hypothesized second r-process event.  

\section{Cosmo chronometry: age dating the oldest stars}\label{chrono}

\begin{figure}[!t]
\begin{center}
\includegraphics[width=13cm, clip=, bb=28 395 525 634]
{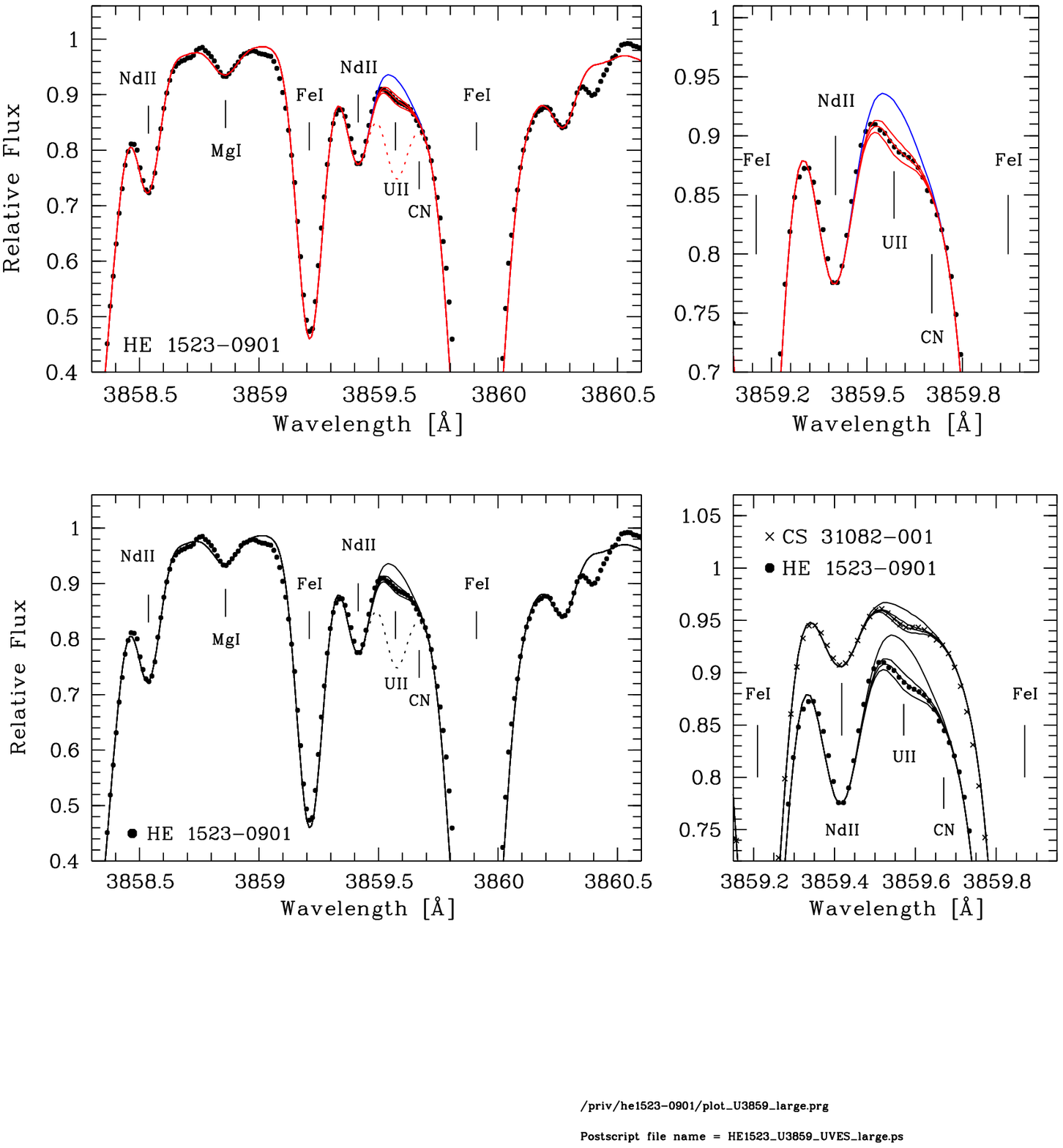}
  \caption{\label{U_region} Spectral region around the U\,II
  line in HE~1523$-$0901 (\textit{filled dots}). The right panel
  is a zoom-in of the region around the U\,II feature.
  Overplotted are synthetic
  spectra with different U abundances. The dotted line in the left
  panel corresponds to a scaled solar r-process U abundance present in
  the star if no U were decayed. Figure adapted from \citet{he1523}.}
\end{center}
\end{figure}

The r-process is responsible for the production of the heaviest
elements, including thorium and uranium. These elements are
radioactive and have long-lived isotopes, $^{232}$Th and $^{238}$U,
with half-lifes of $14$\,Gyr and $4.5$\,Gyr, respectively.  These
half-lives cover cosmic timescales which makes these elements suitable
for age measurements, if found in any objects. Indeed, Th and U can be
detected in some r-process enhanced metal-poor stars if their level of
r-process enhancement is strong enough. Absorption lines of Th are
regularly measured but a U detection is very difficult because only
one, extremely weak, line is available in the optical spectrum.
Figure~\ref{U_region} shows the spectral region around this U line at
3859\,{\AA} in the red giant star HE~1523$-$0901 \citep{he1523}.

Three types of element combinations involving radioactive and
naturally occurring stable r-process elements provide chronometers for
age measurements in metal-poor stars with r-process enhancement that
follow the scaled solar r-process pattern. Examples for stable
elements are europium, osmium and iridium. They are abbreviated with
``r'' in the following equations.

\begin{itemize}
\item[] \mbox{$\Delta t = 46.7[\log{\rm (Th/r)_{initial}} - {\rm
      \log\epsilon(Th/r)_{now}}]$}
\item[] \mbox{$\Delta t = 14.8[\log{\rm (U/r)_{initial}} - {\rm
      \log\epsilon(U/r)_{now}}]$}
\item[] \mbox{$\Delta t = 21.8[\log{\rm (U/Th)_{initial}} - {\rm
      \log\epsilon(U/Th)_{now}}]$}
\end{itemize}

Here, the subscript ``initial'' refers to the theoretically derived
initial production ratio of these elements, while the subscript
``now'' refers to the observed value of the abundance ratio.

The Th/Eu chronometer can regularly be employed and about a dozen
stellar ages of r-process metal-poor stars have been derived this
way. The ages range from $\sim11$ to 14 billion years which make these
star some of the oldest objects in the universe. The U/Th chronometer
was first measured in CS~31082-001 \citep{Hilletal:2002} yielding an age of
$14$\,Gyr.  However, should be noted, that CS~31082-001 suffers from
what has been termed an ``actinide boost'' \citep{honda04}. Compared
with the scaled solar r-process it contains too much Th and U. Hence,
its Th/Eu ratio yields a negative age. The origin of this issue has
yet to be understood. As a result, however, it has become clear that
the r-process material in this and other actinide boost stars likely
have a different origin than other r-process enhanced metal-poor
stars.

Regardless, compared to Th/Eu, the Th/U ratio is  more robust to
uncertainties in the theoretically derived production ratio due to the
similar atomic masses of Th and U \citep{schatz_chronometers}. Hence, stars
displaying Th \textit{and} U are the most valuable old stars.  The U
measurement in HE~1523$-$0901 is the currently most reliable one of
the only three stars with such detections. The availability of both
the Th and U measurements opened up the possibility for the first time
to use seven different chronometers, rather than just one (i.e., Th/Eu
or U/Th). The averaged stellar age of HE~1523$-$0901 derived from
seven abundance ratios involving combinations of Eu, Os, Ir, Th and U
is $13.2$\,Gyr. Realistic age uncertainties, however, range from
$\sim2$ to $\sim5$\,Gyr \citep{he1523}.  

Through individual age measurements, r-process objects become vital
probes for observational ``near-field'' cosmology. Importantly, it
also confirms that metal-poor stars with similarly low Fe abundances
and no excess in neutron-capture elements are similarly old, and that
the commonly made assumption about the low mass (0.6 to
0.8\,M$_{\odot}$) of these survivors is well justified.  Finally,
these stellar ages provide a lower limit to the age of the galaxy and
hence, the universe which is currently assumed to be 13.8\,Gyr
\citep{planck2013}.

\section{Neutron-capture chemical evolution of the galaxy}\label{ncaps}

Up to this point, we have reviewed the neutron-capture abundances of
stars showing large enhancements of neutron-capture species that
reflect (by now) well-established nucleosynthetic patterns.  The vast
majority of metal-poor stars in our galaxy do not show such abundance
enhancements or clear patterns, but inherited small amounts of
neutron-capture material via ``standard'' chemical evolution
(i.e. different processes providing a mix of neutron-capture elements
to star forming gas) over subsequent generations of star formation.

Hence, a natural question to ask when considering neutron-capture
abundances in metal-poor stars is whether some type of
neutron-capture enrichment is seen in all such stars.  The answer
appears to be yes.  Recently, \citet{roederer_srba} carried out a
literature survey of some $\sim$1400 Milky Way and nearby dwarf galaxy
stars, including the most metal-poor (776 stars with
$\mbox{[Fe/H]}\leq-2.0$) known.  Based on a calculation of the
detectability of the strongest Sr and Ba lines available in the
optical regime, he found that every star studied to date has an
abundance measurement or upper limit above the minimum detectable
threshold, indicating ``that no metal-poor stars have yet been found
with sufficiently low limits on [Sr/H] or [Ba/H] to suggest their
birth environment had not been enriched by elements heavier than the
iron group. (p.\ 5)''

\begin{figure}[!ht]
  \begin{center}
     \includegraphics[width=14cm]{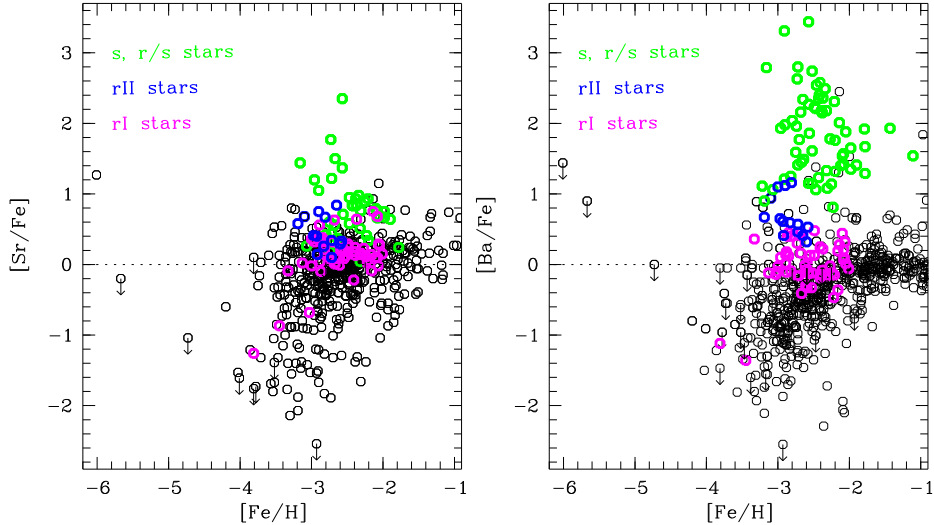}
     \caption{ [Sr/Fe] and [Ba/Fe] versus [Fe/H] for metal-poor stars
       from the catalog of Frebel (2010).  R-process enriched stars
       (where r-I and r-II depict different levels on enhancement),
       s-process enriched stars, and stars showing both r- and s-
       enrichment are indicated by colored points.  Note the degree of
       scatter in neutron-capture element abundances even in ``normal"
       metal-poor stars.  (Figure adapted from
       \citealt{frebel_bookchapter}.) \label{abund} }
    \end{center}     
   \end{figure}

In this section, we thus consider the neutron-capture abundance
patterns in large samples of metal-poor stars that have arisen as a
result of (integrated) galactic chemical evolution and what these
global signatures can reveal about the nature of the r-process(es),
the s-process, and the competition between r- and s-process element
chemical enrichment as a function of time. Figure~\ref{abund} shows
the [Sr/Fe] and [Ba/Fe] abundance ratios versus [Fe/H] for a large
sample of metal-poor halo stars.  The r-process enhanced stars
discussed in Section~\ref{rproc_abund} and the s-process enhanced stars
discussed in Section~\ref{sproc_abund} are indicated by colored symbols; stars
that exhibit no substantial neutron-capture element enhancements or
patterns are shown in black. As can be seen, the dispersion in [Sr/Fe]
and [Ba/Fe] ratios in the ``regular'' metal-poor stars increases with
decreasing [Fe/H] \citep{francois2007, yong13_III, aoki_2013}.

\subsection{Evolution of r-process enrichment}\label{rproc}

Since the discovery of metal-poor stars with strong r-process
enhancement it has been posited that the r-process must have taken
place very early on in the universe. Given that only massive stars
contributed to the element production at that time (the low-mass stars
had not yet evolved enough to significantly contribute to chemical
enrichment) it was proposed that neutrino-driven winds emerging from
the proto neutron star during a supernova explosion would be the natural
site of the r-process (e.g., \citealt{wanajo01, wanajo2002}).  The
massive-star-site also explains, at least qualitatively, the existence
of neutron-capture material in early gas clouds leading to the
formation of metal-poor stars with metallicities of
$\mbox{[Fe/H]}\lesssim-2.7$. From this metallicity upward, the
s-process operating in AGB stars overtakes the global production of
neutron-capture elements and subsequent enrichment of the interstellar
medium (see Section~\ref{sproc} for more details).  

Consideration of the r-process as a main source for the early
neutron-capture history is also important in the interpretation
 of the extremely large scatter observed, for example in
strontium and barium, as shown in Figure~\ref{abund}. At
$\mbox{[Fe/H]}\lesssim-3.0$, there are more than three orders of
magnitude scatter in the [Sr/Fe] and [Ba/Fe] abundance ratios,
pointing to a large variation in neutron-capture abundances while hardly
any variation in iron and presumably other lighter element abundances.  The
scatter is much beyond any measurement uncertainties which could account
for variations on the 0.2\,dex level. Moreover, there appears to be a
systematic trend of Sr being produced more than Ba, as also apparent from
the figure.

Attempts to explain these abundance trends as well as the level of
scatter have been numerous, but with few successes. Generally, a
consensus is emerging that an additional process or processes may have
been at work in the early galaxy, producing preferentially lighter
(e.g., Sr) over heavier neutron-capture elements (e.g.,
Ba). Accordingly, a ``lighter element primary process'' (LEPP) was
proposed \citep{travaglio} to explain the Sr/Ba ratios
observed in metal-poor stars as well as to account for the portion of 
the solar system abundance pattern between Sr and Ba that is not
fully explained by known s-process mechanisms.
A contribution by such a process would
perhaps also be able to explain, to some extent, the deviations from
the scaled solar r-process pattern among light neutron-capture
elements in the strongly r-process enhanced metal-poor stars 
(Section~\ref{rproc_abund}). Other
suggestions include a ``weak'', truncated r-process operating in
$10-20$\,M$_{\odot}$ stars. Here, the entropies and electron fractions
reached in the neutrino-driven wind from the proto-neutrons star are
insufficient to run the r-process to completion, resulting in the preferential
build-up for the lighter r-process species
\citep{wanajo05}.

An alternative explanation of the overall scatter in neutron-capture
element abundances at the lowest metallicities (and hence, earliest
times) is that rather than there being multiple formation sites, there
is only one site in which the r-process reaches various levels of
completion, an extension of the ``weak" r-process scenario above. 
For example, using supernova
nucleosynthesis calculations \citep{woosleyetal1994}, \citet{boyd2012}
modeled a massive ($8-40$\,M$_{\odot}$) star undergoing core collapse
to a neutron star, with r-process reactions going on in the neutrino
driven wind.  These reactions may cease abruptly if the neutron star
collapses into a black hole, taking some of the r-process element
forming regions with it. 

The abundance pattern of neutron-capture elements of the metal-poor
star HD 122563 \citep{honda06} shows a clear gradient of decreasing
abundance with increasing atomic number. This could perhaps be the
pattern emerging from this suspected process that produces light
neutron-capture elements over heavier ones.  Such a truncated
r-process indeed provides a reasonable fit to the observed abundances of this heavy
neutron-capture element depleted star.

To explain the huge scatter in neutron-capture elements,
\citet{aoki_2013} incorporated the truncated r-process scenario of
Boyd et al.\ in the galactic chemical evolution model of
\citet{cescutti2006}.  They found they were able to reproduce the
range of [Sr/Ba] ratios found in metal-poor stars.  This scenario may
also explain the very different levels of scatter of neutron-capture
elements compared to $\alpha$-elements, if the production of
$\alpha$-elements is occuring in layers further out from the
proto-neutron star and can therefore more easily escape collapse onto
the black hole. Overall, chemical evolution models that incorporate
multiple (at least two) r-process sites are better able to reproduce
the abundance patterns of metal-poor stars
\citep{qian_wasserburg2007}. To complicate matters, it is likely that
the s-process operating in fast-rotating massive stars (see
Section~\ref{sproc}) may also contribute to the scatter seen at the
lowest metallicities. With more observational data and improved
theoretical modeling of the process and sites, the contributions of
all these processes can be disentangled.

\subsection{Evolution of s-process enrichment}\label{sproc}

The onset of s-process enrichment in
our Galaxy, that is the ``time" at which the first intermediate- and
low-mass stars had reached the AGB phase and began to pollute their
environments with s-process material, has been the subject of careful
study.  There is some uncertainty in the literature over when this
process begins.  A study of the relative abundances of Ba and Eu by
\citet{2000burris} showed some indication that s-process enrichment
set in by $-2.4<\mbox{[Fe/H]}<-2.1$, but in some stars may appear as
early as $\mbox{[Fe/H]}\sim-2.75$ ([Fe/H] being a ``chemical time".)
\citet{simmerer2004} found signs of its presence at
$\mbox{[Fe/H]}\sim-2.6$, but remarked on the large spread in s-process
element [X/Fe] ratios, even at solar [Fe/H] in their stellar sample.
Similarly, \citet{hansen13_agpd} found evidence of the s-process at
$\mbox{[Fe/H]}\sim-2.5$, due to flattening of abundance trends with
[Fe/H] and general decrease in scatter at that point.  However, other
studies have found first signs of the s-process enrichment much later,
at $\mbox{[Fe/H]}\sim-1.5$ \citep{roederer_ubiq, zhang_holistic}.

The short lifetimes of rapidly-rotating massive metal-poor stars
(e.g., \citealt{meynet06}, \citealt{hirschi07}) makes these stars
potentially significant contributors to the s-process production in
the early universe (recall Section~\ref{sproc_nucleo}) before the
onset of s-process enrichment through low-mass AGB stars. These
spinstar models predict that large amounts of scatter in
neutron-capture element-to-iron ratios, as well as in ratios of light
(e.g., Sr) to heavy (Ba) element abundances, which is what is seen in
metal-poor stars (Figure~\ref{abund}). Chemical evolution models of
the galactic halo that include yields of spinstars along with chemical
enrichment from $8-10$\,M$_{\odot}$ stars via the main r-process
reproduce the scatter in neutron-capture element abundance trends in
metal-poor stars very well \citep{chiappini11, cescutti_spin}.
Furthermore, \citet{cescutti_spin} posit that the s-process
contribution of spinstars in the early universe also provide a natural
explanation for why no metal-poor star has yet been observed to lack
neutron-capture element abundances (\citealt{roederer_srba}).

\section{Summary and open questions}

We have presented an overview of neutron-capture abundance studies of
metal-poor stars and how the different groups of stars can help to
constrain either individual nucleosynthesis processes such as the r-
and the s-process, their sites of operation, and the overall process
of chemical evolution that is driven by a variety of sites with time.
 
The wealth of data on the different neutron-capture processes not only
helps to reconstruct the formation and evolution of the heaviest
elements, but also opens new questions for which answers have not yet
been found. For example, the deviations from the scaled solar
r-process pattern remain puzzling both observationally and
theoretically. Then, details of the mass transfer events aross binary
systems and the subsequent dilution processes of s-process material
into the atmosphere of the metal-poor stars still observable today are
not well understood. However, our interpretations regarding the nature
of the s-process at low-metallicity depend in part on knowledge of
this process.

Another interesting observational finding is that all the strongly
r-process enhanced metal-poor stars found so far exhibit a narrow
range in [Fe/H] of $0.3-0.4$\,dex. If the r-process is universal, why
do these stars appear at a certain ``chemical time'', as put by \citet{hansen2011}?
  Some propose this signals the start of a new process
at work in the chemical evolution of the universe (e.g., \citealt{francois2007}), 
or else that these stars only form from a very special
type of supernova in which the neutron-capture elements are released
via jets, unlike the other elements \citep{hansen2011}. More generally, no stars with
$\mbox{[Fe/H]}<-3.5$ have yet been discovered that display any known
or charactistic neutron-capture abundance pattern. This raises the
question of when exactly the very first neutron-capture events took
place in the early universe and whether massive Population\,III stars
produced neutron-capture material, and if so, in what
quantities. Finally, abundances of neutron-capture elements with
$40<Z<56$, i.e. those between the first and second peak, signal that
yet other, unidentified neutron-capture processes may have been at
work in the early universe.  In their analysis of silver and palladium in
metal-poor stars, \citet{hansen13_agpd} found that the abundance ratios
of Pd and Ag (e.g., [Ag/Fe], [Ag/Eu], [Ag/Ba]) did not match the patterns expected if they
were produced by the main r, the weak r, or any s-process channel.

Only with more and better data of existing and to be discovered
metal-poor stars will some of these questions be answered. Upcoming
telescope projects such as the 25\,m Giant Magellan Telescope, when
equipped with high-resolution spectrographs, in combination with
targets selected from large-scale sky surveys will greatly advance the
discovery process of many extremely metal-poor stars to move
observational nuclear astrophysics into a new era.

\vspace{10 mm}
We thank Marco Pignatari for helpful comments and suggestions on parts of
the manuscript.  Amanda Karakas is also gratefully acknowledged for
making Figure~\ref{sproc_pattern} for this review, as well as for her
comments on the final draft.  



\end{document}